\documentclass[prl,twocolumn,showpacs,amsmath,amssymb]{revtex4}
\usepackage{graphicx}
\usepackage{dcolumn}
\usepackage{bm}

\begin{document}

\title{Optimal error tracking via quantum coding and continuous syndrome measurement}
\author{Ramon van Handel} \email{ramon@caltech.edu}
\author{Hideo Mabuchi} \email{hmabuchi@caltech.edu}
\affiliation{Physical Measurement and Control 266-33, California
Institute of Technology, Pasadena, CA 91125}

\date{\today}

\begin{abstract}
We revisit a scenario of continuous quantum error detection proposed
by Ahn, Doherty and Landahl [Phys.\ Rev.\ A {\bf 65}, 042301 (2002)]
and construct optimal filters for tracking accumulative errors.
These filters turn out to be of a canonical form from hybrid control
theory; we numerically assess their performance for the bit-flip and
five-qubit codes. We show that a tight upper bound on the stochastic
decay of encoded fidelity can be computed from the measurement
records. Our results provide an informative case study in
decoherence suppression with finite-strength measurement.
\end{abstract}

\maketitle

\noindent Prospects for quantum computation have motivated the
development of an extensive theory of quantum error
prevention/correction (QEC) \cite{Ali05}. Despite experimental
demonstrations of some key methods \cite{Lan05,Chi04,Kni01}, there
remains a significant gap between the abstract algebraic theory of
QEC and concrete physical models of real quantum systems. In
particular, QEC is still mainly discussed in terms of instantaneous
measurement and recovery operations rather than more realistic
continuous-time dynamical models. Bridging this gap would enable us,
{\it e.g.}, to deduce limits on QEC performance from the finite
speed of laboratory measurement and recovery operations. Appropriate
continuous-time formulations of QEC would likewise facilitate deeper
connections with control theory, which could in turn lead to the
discovery of new quantum memory schemes.

Our aim in this article will be to pursue a rigorous approach to
incorporating finite-strength measurement within the familiar
conceptual setting of discrete quantum error correcting codes. At
the same time, we will arrive at familiar equations from classical
control theory that should be amenable to characterization via
established methods of stochastic analysis. While the specific
models we consider---continuous-time relaxations of stabilizer codes
as introduced by Ahn, Doherty and Landahl \cite{Ahn}---are rather
impractical from an experimental point of view, they do provide a
canonical setting in which to demonstrate how essential ideas from
quantum error correction can be reconciled with stochastic control
theory. We regard this as an important step towards developing a
general theory of optimal decoherence suppression with resource
constraints on measurement and control.

We begin by considering a continuous-time relaxation of the bit-flip
code \cite{Ahn}. A logical qubit state $|\Psi\rangle =
c_0|0\rangle+c_1|1\rangle$ is encoded in the joint state of three
physical qubits as $|\Psi\rangle\mapsto
c_0|000\rangle+c_1|111\rangle \equiv |\Psi_E\rangle$. The system is
coupled to decay channels that can cause independent Markovian bit
flips and also to probe channels that allow for continuous
(finite-strength) syndrome measurement \cite{Saro04}. Following
\cite{Ahn,Saro04}, we assume that the syndrome measurements are
constructed by coupling two probe fields to the syndrome generators
\cite{Gottesman} $M_1=ZZI=\sigma_z\otimes\sigma_z\otimes\openone$
and $M_2=ZIZ=\sigma_z\otimes\openone\otimes\sigma_z$. The total
system dynamics is then described by the quantum stochastic
differential equation (e.g.\ \cite{GaZo})
\begin{multline}
\label{eq:QSDE}
    dU_t=\{\sqrt{\gamma}\,(X_1\,dB^{1\dag}_t+X_2\,dB^{2\dag}_t+
        X_3\,dB^{3\dag}_t-\mbox{h.c.})+\\
    \sqrt{\kappa}\,(M_1\,dA^{1\dag}_t+M_2\,dA^{2\dag}_t-\mbox{h.c.})
    -(\tfrac{3}{2}\gamma+\kappa)\,dt\}\,U_t
\end{multline}
where $B^i_t$ are the bit-flip channels, $A^i_t$ are the probe
channels and $X_1=XII=\sigma_x\otimes\openone\otimes\openone$,
$X_2=IXI$ {\it et cetera}. The unitary evolution $U_t$ is a
Schr{\"o}dinger picture propagator for the entire system including
the probe fields.

For homodyne-type detection of the two probe channels we obtain two
observation processes $Y^i_t=U_t^\dag(A^i_t+A^{i\dag}_t)U_t$ that
can be written (via quantum It\^o rules) as
\begin{equation}
    dY^i_t=2\sqrt{\kappa}\,U_t^\dag M_i U_t\,dt + dA^i_t+dA^{i\dag}_t
\end{equation}
(this is called the input-output picture in \cite{GaZo}). Similarly,
we could in principle attempt to observe the bit-flip events
directly by performing direct detection of the corresponding decay
channels $B_t^i$; this would give rise to additional (counting)
observations $Z^i_t$ that are independent Poisson processes with
rate $\gamma$ (see, {\it e.g.}, \cite{BvH05}). Although the $Z_t^i$
are generally assumed to be unobservable in QEC, we will exploit
them below to obtain useful information on the statistics of the
measurement currents $Y_t^i$.

Our goal is to detect and to correct bit-flip errors by making use
of the probe currents $Y_t^i$, $i=1,2$. Hence we must understand how
best to extract information about error events from the noisy
observed signals. We begin by calculating the least mean-square
estimator for our system \cite{BvH05}, in the form of a (random)
three-qubit density matrix $\rho_t$ (the conditional state). For
every system observable $A$, ${\rm Tr}[A\rho_t]$ is the function of
the observation history that minimizes the estimation error $\langle
(U_t^\dag AU_t-{\rm Tr}[A\rho_t])^2\rangle$. If we observe only
$Y^i_t$, then $\rho_t$ obeys the quantum filtering equation
\cite{BvH05}
\begin{multline}
\label{eq:filter}
    d\rho_t=
    \sum_{k=1}^3\gamma\,\mathcal{T}[X_k]\rho_t\,dt
    +\sum_{i=1}^2\kappa\,\mathcal{T}[M_i]\rho_t\,dt \\
    +\sum_{i=1}^2\sqrt{\kappa}\,
    \mathcal{H}[M_i]\rho_t\,(dY_t^i-2\sqrt{\kappa}\,
        {\rm Tr}[M_i\rho_t]\,dt)
\end{multline}
where $\mathcal{T}[X]\rho=X\rho X^\dag-\rho$ and
$\mathcal{H}[X]\rho= X\rho+\rho X^\dag-{\rm
Tr}[(X+X^\dag)\rho]\rho$. This is equivalent to the stochastic
master equation used in \cite{Ahn}. If we were to additionally
observe the bit flip channels $Z_t^i$, we could obtain an improved
estimator $\tilde\rho_t$ that would obey a ``jump-unraveled''
version of Eq.~(\ref{eq:filter}) with $\gamma\mathcal{T}[X_k]\,dt$
replaced by $\mathcal{T}[X_k]\,dZ_t^i$.
An important feature of these equations \cite{BvH05} is that the
so-called innovations processes $dW_t^i=dY_t^i-2\sqrt{\kappa}\,{\rm
Tr}[M_i\rho_t]\,dt$ are independent and have the law of a Wiener
process.

We can use the fact that the innovations are Wiener processes, in
the absence of `real' (physically generated) measurement signals
$Y_t^i$, to generate the latter through Monte Carlo simulations. By
driving Eq.~(\ref{eq:filter}) with random sample paths of a Wiener
process, we can reconstruct observation processes $Y_t^i$ that
sample the space of measurement records with the correct probability
measure. The evolution of the filter variables in each simulation
fairly represents what would have happened if physically generated
measurement records $Y_t^i$ had actually been presented to the
filter and it derived $W_t^i$ from them. Similarly, we can
reconstruct $Y_t^i$ from the jump-unraveled version of
Eq.~(\ref{eq:filter}) by driving it with Wiener processes $\tilde
W_t^i$ and independent Poisson processes $Z_t^i$ with rate $\gamma$.
The innovations theorem guarantees that both simulations will
generate sample paths $Y_t^i$ with the same probability measure. We
will invoke this property later.

In the usual setting of discrete quantum codes, an instantaneous
measurement of $M_1$ and $M_2$ after a period of free evolution is
used to determine the recovery operation (if any) that should be
applied. If $(M_1,M_2)=(+1,+1)$ no correction is necessary; outcomes
$(-1,+1)$ mean that $IXI$ should be applied; $(+1,-1)\Rightarrow
IIX$ and $(-1,-1)\Rightarrow XII$. These outcomes are called the
error syndromes. In the continuous setting we introduce the
orthogonal projectors $\Pi_0\ldots\Pi_3$ onto the eigenspaces
corresponding to each syndrome. The quantity $p^m_t={\rm
Tr}[\Pi_m\rho_t]$ is then the conditional probability (given the
noisy probe observations) that, had we actually measured $M_1$ and
$M_2$, we would have obtained the syndrome corresponding to $\Pi_m$.
Plugging $p^m_t$ into Eq.~(\ref{eq:filter}) we obtain the syndrome
filter
\begin{equation}
\label{eq:wonham}
    dp_t = \Lambda^T p_t\,dt+
    \sum_{i=1}^2 (H_i-h_i^Tp_t\,\openone)p_t\,(dY_t^i-h_i^Tp_t\,dt)
\end{equation}
where $h_i^m/2\sqrt{\kappa}$ is the outcome of $M_i$ corresponding
to the syndrome $\Pi_m$, $H_i=\mbox{diag}\,h_i$, and
$\Lambda_{mn}=\gamma(1-4\delta_{mn})$. The $p_t^i$ form a closed set
of equations, as the observations are uninformative on the logical
state of the qubit and the coherences (if any) between the
syndromes.

\begin{figure}
\includegraphics[width=0.45\textwidth]{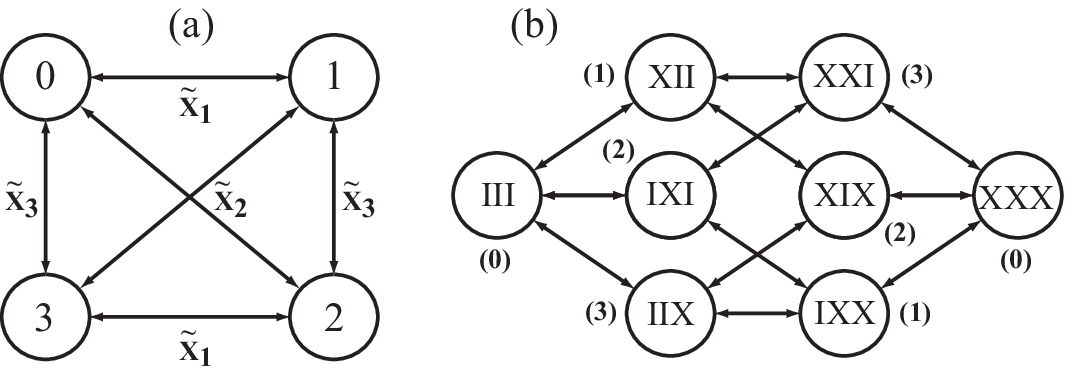}

\noindent (c)

\includegraphics[width=0.45\textwidth]{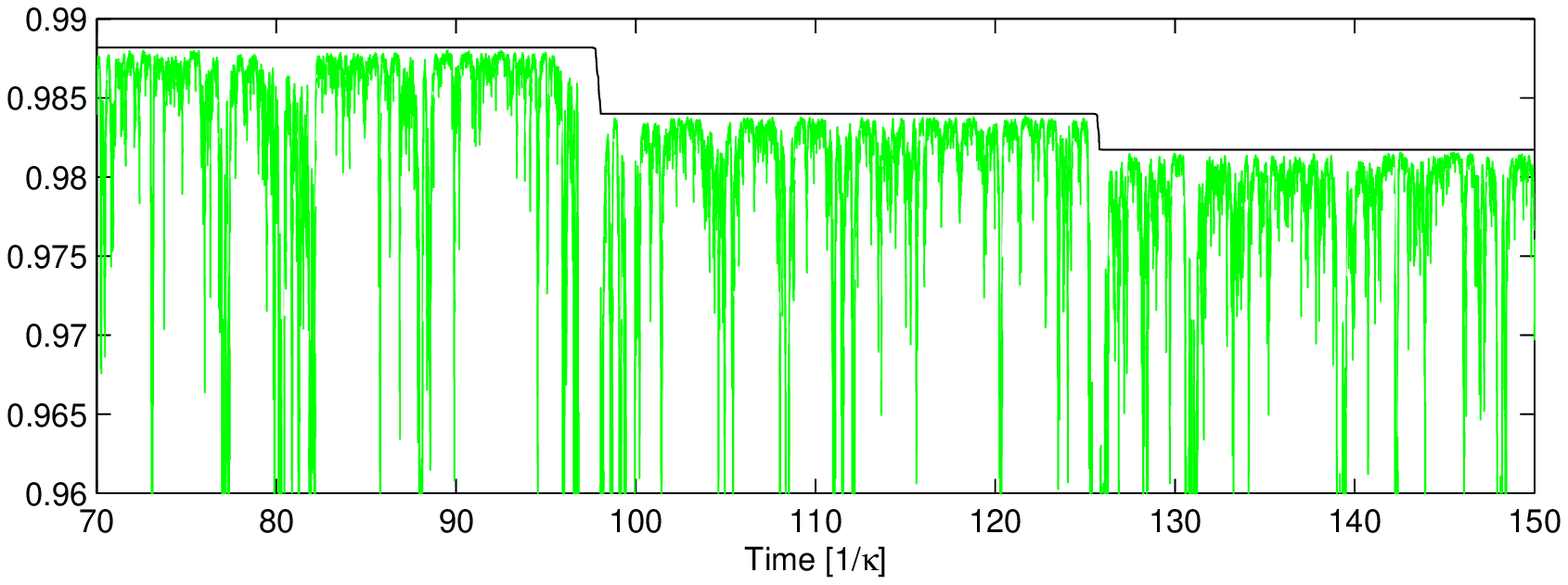}
\caption{\label{fig:syn}
    Markov chains associated to the three-qubit code:
    (a) the syndrome chain, and (b) the extended correction chain
    with the corresponding syndromes labeled between brackets.
    All transitions are independent with rate $\gamma$.
    (c) A Monte Carlo trajectory of $\max_i(p_t^i)$ (green noisy
    curve) and $\mathfrak{J}_t$ (black step-like curve) for the
    five-qubit code with $\kappa/\Gamma=100$.}
\vspace{-0.1in}\end{figure}

The syndrome filter, Eq.~(\ref{eq:wonham}), is a familiar equation
from classical probability theory; we will gain important insight by
reducing our problem to a classical one. Consider a system that can
be in one of four states labeled $0\ldots 3$. Suppose the system is
in some known state at time $t$. After an infinitesimal time
increment $dt$ the system can switch to one of the other three
states, each of which occurs with probability $\gamma\,dt$. This
defines a Markov jump process on a graph, as depicted in
Fig.~\ref{fig:syn}a. Unfortunately we cannot observe the state
directly; instead, we are given two observation processes of the
form $dY_t^i=h_i^{m_t}\,dt+dV_t^i$ where $V_t^{1,2}$ are two
independent Wiener processes that corrupt our observations, and
$m_t$ is the state of the Markov jump process at time $t$. As our
observations are noisy we cannot know the system state with
certainty at any time. However, we can calculate the {\it
conditional} probability $p_t^m$ that it is in state $m$ at time
$t$. This classical estimation problem is precisely solved by
Eq.~(\ref{eq:wonham}), known as the Wonham filter \cite{Wonham,EAM}.

To help interpret this result, consider the jump-unraveled version
of Eq.~(\ref{eq:filter}). In the same way that we obtained
Eq.~(\ref{eq:wonham}), we can substitute $\tilde p_t^m={\rm
Tr}[\Pi_m\tilde\rho_t]$ and get a closed form expression. Assuming
the initial state lies inside one of the syndrome spaces \footnote{
    This is not an essential restriction, as the probability measure
    on the space of measurement records can be written for any initial
    state as the corresponding mixture of such measures given a fixed
    initial syndrome.
}, it is readily verified that $[M_1,\tilde\rho_t]=
[M_2,\tilde\rho_t]=0$ for all $t$ and we obtain
\begin{equation}
    d\tilde p_t=\sum_{k=1}^3(\tilde X_k-\openone)\tilde p_t\,dZ_t^k
\end{equation}
where $\tilde p_0$ is one of the unit vectors $e_n^m=\delta_{nm}$.
Here $\tilde X_1$ is a matrix such that $\tilde X_1e_0=e_1$, $\tilde
X_1e_1=e_0$, $\tilde X_1e_2=e_3$, and $\tilde X_1e_3=e_2$, and
$X_{2,3}$ are defined similarly as shown in Fig.~\ref{fig:syn}a. The
solution of this equation is of the form $\tilde p_t=e_{m_t}$ where
$m_t$ is the Markov jump process defined above. Since
\begin{equation}
    d\tilde W_t^i=dY_t^i-h_i^T\tilde p_t\,dt=dY_t^i-h_i^{m_t}\,dt
\end{equation}
must be a Wiener process that is independent from all $Z_t^i$, the
statistics of the probe observations obtained from the quantum
system are precisely described by the classical model of the
previous paragraph. The Markov process $m_i$ is simply the error
syndrome obtained by observing the bit flips directly, and the
Wonham filter above has a natural classical interpretation as the
best estimate of $m_i$ given only the noisy probe observations.

We now turn to the problem of error correction. Suppose that we let
the system evolve for some time while propagating the filter
Eq.~(\ref{eq:wonham}) with the observations. At some time $T$ we
pose the question: what operation, if any, should we perform on the
system to maximize the probability of restoring the initial logical
state $|\Psi_E\rangle$? We will assume that we can pulse the system
sufficiently strongly (as compared to $\kappa,\gamma$) to perform
essentially instantaneous bit flips on any of the physical qubits.
The most obvious decision strategy simply mimics discrete error
correction---given the most likely syndrome state
$m_*=\mbox{arg\,max}_m\,p_T^m$, we do nothing if $m_*=0$ and
otherwise we perform a bit flip on physical qubit $m_*$.

But it is possible to do better. Recall that the discrete error
correction strategy is based on an assumption that at most one bit
flip occurs in the interval $[0,T]$. This assumption may not hold in
practice. With our continuous syndrome measurement, we do actually
have some basis for estimating the total number (and kind) of bit
flips that have occurred. Unfortunately this information does not
reside in the statistic $p_t$, which only gives the conditional
probabilities of the syndromes at the current time. We are seeking a
non-Markovian decision policy that knows something about the history
of the bit flips.

The classical machinery introduced above allows us to solve this
problem {\it optimally}. To do this we simply extend the Markov jump
process $m_t$ as shown in Fig.~\ref{fig:syn}b. The states of the
extended chain $\hat m_t$ are no longer the four syndromes but the
eight {\em error states} that may obtain at any given time. Every
syndrome corresponds to two error states, as is shown in
Fig.~\ref{fig:syn}b. We still consider the same observation
processes, so error states that belong to the same syndrome give
rise to identical observations. Thus on the basis of the
observations, the two Markov chains are indistinguishable.
Nonetheless the extended chain gives rise to a different estimator
that provides precisely the information we want. As by construction
the Wonham filter provides the optimal estimate, we conclude that
the optimal solution to our problem is given by the Wonham filter
for the extended chain, {\it i.e.}, the eight-dimensional equation
\begin{equation}
\label{eq:wonham2}
    d\hat p_t = \hat\Lambda^T\hat p_t\,dt+
    \sum_{i=1}^2 (\hat H_i-\hat h_i^T\hat p_t\,\openone)
        \hat p_t\,(dY_t^i-\hat h_i^T\hat p_t\,dt)
\end{equation}
where $\hat\Lambda$, $\hat h_i$ are the intensity matrix and
observation vector for the chain $\hat m_t$ (see \cite{Wonham,EAM}
for details). The optimal correction policy is now simple: at time
$T$, we perform the correction that corresponds to the state $\hat
m_*=\mbox{arg\,max}_m\,\hat p^m_T$. Hence if $\hat m_*=III$ we do
nothing, if $\hat m_*=IXX$ we flip physical qubits 2 and 3, etc.
This maximizes the probability of restoring $|\Psi_E\rangle$.

From Fig.~\ref{fig:syn}b we can see how information is lost from the
quantum memory. At time $t=0$ we begin in the no-error state $III$.
A bit flip might occur which puts us, {\it e.g.}, in the state
$XII$, then $XIX$, {\it etc}. But as we are observing these changes
in white noise there is always a chance that when two bit flips
happen in rapid succession, we ascribe the corresponding
observations to a fluctuation in the white noise background rather
than to the occurrence of two bit flips. In essence, successive bit
flips are resolvable only if they are separated by a sufficiently
long interval that the filter can average away the white noise
fluctuations (the signal-to-noise $4\kappa$ determines this
timescale). Occasionally, multiple bit flips occur too rapidly and
information is lost ({\it e.g.}, $III\rightarrow XII\rightarrow XIX$
may be mistaken for $III\rightarrow IXI$ since the two final states
have identical syndromes). It is evident from Fig.~\ref{fig:syn}b
that this rate of information loss is independent of the error
state. Hence there is no point in applying corrective bit flips at
intermediate times $t<T$, as this cannot slow the loss of
information. As the Wonham filter is optimal by construction, we
conclude that the correction policy described above is optimal
\footnote{
    This assumes that we trust Eq.~(\ref{eq:QSDE}) completely. If
    there is some uncertainty in the model it is sometimes
    advantageous to consider different estimators \cite{James}.
}.

How can we quantify the information loss from the system? By
construction $\hat p_t^*=\max_m\hat p_t^m$ is the probability of
correct recovery at time $t$. Unfortunately, as one can see in
Fig.~\ref{fig:syn}c, the quantity $\hat p_t^*$ fluctuates rather
wildly in time. Thus it is not a good measure of the information
content of the system, as it is very sensitive to the whims of the
filter: the filter may respond to fluctuations in the measurement
record by adjusting the state as if a bit flip had occurred, but
then correct itself when it becomes evident that nothing happened.
We actually want to find some quantity that gives a (sharp) upper
bound on all {\em future} values of $\hat p_t^*$. This would truly
measure the information content of the system, as it bounds the
probability of correct recovery that can be achieved.

We claim that the quantity $\mathfrak{J}_t=\max_m(\hat p_t^m/(\hat
p_t^m+\hat p_t^{\bar m}))$, which is a function of filter variables,
provides a suitable measure of the information content at time $t$.
Here $\bar m\ne m$ is the error state that corresponds to the same
syndrome as $m$, so $\hat p_t^m+\hat p_t^{\bar m}=P_t^m$ is the
probability of the syndrome corresponding to $m$. Hence we can
interpret $\mathfrak{J}_t$ as the conditional probability of the
error state $m$, given that the system is in the corresponding
syndrome. Define $I_t^m=\hat p_t^m/(\hat p_t^m+\hat p_t^{\bar m})$
so that $\mathfrak{J}_t=\max_mI_t^m$. Direct application of the
It\^o rules gives
\begin{equation}
    \frac{dI_t^m}{dt}=-\sum_{n\ne m}\hat\Lambda_{nm}\,
    \frac{P_t^n}{P_t^m}\,(I_t^m-I_t^n).
\end{equation}
If we define $m_t^*=\mbox{arg\,max}_m\,I_t^m$, then we get
\footnote{
    To make the argument completely rigorous one must check that
    this equation is well defined, i.e.\ that $d\mathfrak{J}_t/dt$
    exists.  This can be done using the methods in \cite{Chigansky}.
}
\begin{equation}
    \frac{d\mathfrak{J}_t}{dt}=-\sum_{n\ne m_t^*}\hat\Lambda_{nm_t^*}\,
    \frac{P_t^n}{P_t^{m_t^*}}\,(\mathfrak{J}_t-I_t^n)
    \le 0.
\end{equation}
Hence $\mathfrak{J}_t$ decreases monotonically, and moreover by
construction we must have $\hat p_t^m\le I_t^m$, so $\hat p_t^*\le
\mathfrak{J}_t\le \mathfrak{J}_s$ for all $s<t$. Thus evidently
$\mathfrak{J}_t$ bounds all future values of $\hat p_t^*$. One would
expect the bound to be tight for sufficiently high signal-to-noise
(as then $P_t^{m_t^*}$ will be close to one), which is indeed the
case as can be seen in Fig.~\ref{fig:syn}c.

\begin{figure}
\includegraphics[width=0.47\textwidth]{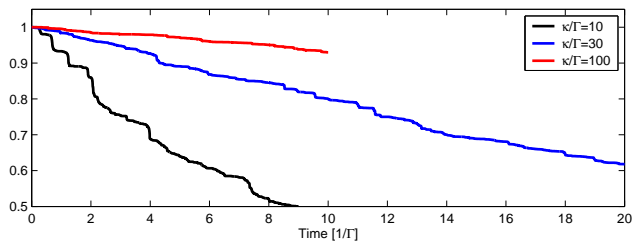}
\caption{\label{fig:perf}
    Average decay of $\mathfrak{J}_t$ for the five-qubit code (dependence
    on $\kappa/\Gamma$). These curves indicate upper bounds on
    the performance of a quantum memory based on coding and
    continuous syndrome measurement.}

\vspace{-0.1in}
\end{figure}

The procedure we have outlined can be generalized to other
stabilizer codes, such as the five-qubit code \cite{Gottesman}. This
code protects one logical qubit against single-qubit errors by
encoding in five physical qubits and measuring four stabilizer
generators. For `Pauli channel' decoherence described by Lindblad
terms $\sum_{k=1}^5\gamma\,(\mathcal{T}[X_k]+\mathcal{T}[Y_k]+
\mathcal{T}[Z_k])$, the error state graph can be constructed by
considering all possible assignments of an error state
$\in\{I,X,Y,Z\}$ to each qubit and by connecting every pair of
states that are related by the action of a Pauli operator
$\in\{X,Y,Z\}$ on one qubit. One thus has a graph with $4^5=1024$
nodes, with each node connected to $3\times 5=15$ other nodes (we
have validated this construction by comparing simulations of the
corresponding Wonham filter with an appropriate stochastic master
equation). The total error rate is $\Gamma=15\gamma$.
Fig.~\ref{fig:syn}c shows a portion of a single Monte Carlo
simulation of the Wonham filter for the five-qubit code;
Fig.~\ref{fig:perf} shows averages of $\mathfrak{J}_t$ over tens of
trajectories each for $\kappa/\Gamma\in\{10,30,100\}$.

In conclusion, we have shown that both the three- and five-qubit
codes are amenable to a classical analysis in terms of Markov jump
processes, which enables an optimal solution of the error tracking
problem in continuous time. Though the filters that must be
propagated for this purpose are high-dimensional, the optimal
procedure gives at least an upper bound on the achievable
performance of quantum memories based on coding and finite-strength
syndrome measurement.
%
In practical situations one might not have sufficient resources to
propagate the full optimal filter, so suboptimal methods are clearly
of interest.

One could also consider different control goals. For example,
suppose that rather than requiring the state to be corrected at time
$T$, we allow ourselves a little more time afterwards. This can be
helpful; if the syndrome has just jumped, but we have not had enough
time to observe this yet, then we can avoid an incorrect recovery by
waiting just long enough to observe the jump. This can backfire,
however, as waiting too long will just cause us to lose information.
The optimal solution to this problem is known as an optimal stopping
problem \cite{Shiryaev} and is studied extensively in the
mathematical finance literature \cite{OksSulem}. All of these
problems, and many others, are subsumed under the title of {\it
hybrid control theory}. It is our hope that this theory will provide
important tools for the analysis and design of continuous quantum
error correction codes and suboptimal estimators, and for the
solution of the associated control problems.

\begin{acknowledgments}
This work was supported by the Army Research Office under Grant
DAAD19-03-1-0073. HM thanks D.\ Poulin and M.\ Nielsen for
insightful discussions.
\end{acknowledgments}

\bibliographystyle{apsrev}
\bibliography{QEC}

\end{document}